\documentclass[letter]{jpsj3}
\usepackage{txfonts}
\usepackage{graphicx}
\usepackage{dcolumn}
\usepackage{bm}
\usepackage{color}

\def\beq{\begin{equation}}
\def\eeq{\end{equation}}
\def\beqa{\begin{eqnarray}}
\def\eeqa{\end{eqnarray}}
\def\adag{a^{\dagger}}
\def\bdag{b^{\dagger}}
\def\fdag{f^{\dagger}}

\def\degC{\kern-.2em\r{}\kern-.3em C}

\title{Real Time Relaxation Dynamics of  
Macroscopically Photo-Excited Electrons 
toward the Fermi Degeneracy Formation  in the Conduction Band of Semiconductors}

\author{Hiromasa Ohnishi$^1$\thanks{hohnishi@tsuruoka-nct.ac.jp}, Norikazu Tomita$^2$, and Keiichiro Nasu$^3$}
\inst{$^1$National Institute of Technology (NIT), Tsuruoka College, 104 Sawada, Inooka, Tsuruoka, 
Yamagata, 997-8511,Japan\\
$^2$Department of Physics, Yamagata University, 1-4-12 Kojirakawa, Yamagata, 
990-8560, Japan\\
$^3$Institute of Materials Structure Science, High Energy Accelerator Research
Organization (KEK), 1-1 Oho, Tsukuba, 305-0801, Japan} 

\abst{Concerning with the recent experiment of 
time-resolved two-photon photo-emission spectral measurements 
on semiconductors (GaAs, InP), 
we theoretically study real time relaxation dynamics of 
macroscopically photo-excited electrons, 
toward the Fermi degeneracy formation in an originally vacant conduction band 
of these semiconductors.
Very soon after the photo-excitation, 
the whole electrons are shown to exhibit a quite rapid relaxation, 
like an avalanching phenomenon, mainly due to successive multi-(optical and acoustic) 
phonon emission from them. 
Repeating this multi-phonon process, 
the whole energy distribution of the electrons is shown to become 
a multi-peaked structure largely elongated over the lower part of the wide conduction band. 
However, after around 1 ps from the excitation, 
this relaxation critically slows down, 
since the emission of a long-wave acoustic phonon from electrons around the Fermi level 
becomes prohibitively difficult. 
By using the electron temperature approximation,
we show that this slow relaxation is inversely proportional to time.
Thus, the formation of the complete Fermi degeneracy takes an infinite time. 
These theoretical results are quite consistent to the aforementioned recent experiment.
}

\begin{document}
\maketitle

Presence of the Fermi surface or Fermi degeneracy is 
one of the most fundamental concepts in the solid state physics.
In fact, one can easily see back the standard textbook, 
that the most of many body theories for solids, 
say the BCS, the Kondo and the CDW theories, 
all start after the Fermi surface has been already 
well established \cite{mahan}. 
In the present paper, however, 
let us return to a more original viewpoint, 
and ask how this Fermi degeneracy itself can be 
generated from the very beginning, 
or from the true electron vacuum, 
to which macroscopic number of electrons are photo-injected.

Generally speaking, 
when a macroscopically condensed system is shone by photons, 
electrons in this system will usually be excited, 
and after a while, the system will relax down to the original ground state. 
Nowadays, we can observe this transient relaxation process of electrons 
in detail as a function of real time, 
through the time-resolved photo-electron emission spectra. 

In metallic, or highly conductive systems, 
rapid relaxation dynamics of optically excited electrons 
has already been well-known and widely investigated \cite{cu,au,zhukov,baranov}. 
In most cases, however, only a minor part of the whole electrons is excited, 
while the main part of electrons is still in the original ground (Fermi degenerate) state and 
works as an infinite heat reservoir, 
resulting in a quite rapid relaxation or dissipation of newly given energy and momentum. 
In typical cases, only a few femtoseconds (fs) after the photo-excitation, 
are enough to return as far as to the vicinity of the Fermi degeneracy \cite{zhukov}.

Now, we may have a quite naive but interesting question. 
What happens if a small but macroscopic 
number of electrons are excited at once into a truly vacant conduction band 
without electronic heat reservoir at low (absolute zero) temperature? 
Even in this case with no electronic heat reservoir, the excited 
electrons will relax down toward the Fermi degeneracy at around the bottom of the conduction band. 

This ideal situation is known to be experimentally realized and can be investigated as a practical phenomenon by using the direct gap semiconductors such as GaAs and InP so far, but the relaxation mechanism of photo-excited carriers is still 
under considerable debates \cite{potz, thomas, rota, collet, kurz, kaiser}.

Very recently, however, a time-resolved two-photon photo-emission spectroscopy measurement 
in GaAs and InP has been performed \cite{kanasaki0,kanasaki}, 
in which we can directly access to 
a transient time evolution of carrier (electron) distributions in the conduction band. 
The essential results of this experiment are as follows: 
(1) The electrons are excited, by an intense visible laser pulse, 
to the states around 0.3eV higher from the conduction band minimum (CBM). 
(2) The Fermi energy is located around 0.1eV above this CBM, 
implying the carrier number is about 0.003 electrons per unit cell. 
(3) In the initial 1 picosecond (ps) from the photo-excitation, 
a rapid shift of the whole electron distribution like an avalanche is observed. 
In this process, a multi-peaked structure of electron distribution is observed.
It is elongated over the lower part of the wide conduction band. 
The important point in this early stage is that such an electron 
distribution is far from the Fermi-Dirac one, and we cannot define the 
electronic temperature. 
(4) After 1 ps from the photo-excitation, 
the change of the electron distribution becomes very small 
and the relaxation keeps slowing down. 
This slow relaxation, in the long time limit, is inversely proportional to time.
Thus, the situation is quite different from 
the aforementioned metallic or highly conductive cases.

In this letter, 
we theoretically study how a macroscopic number of photo-excited electrons will relax 
toward the Fermi degeneracy formation in the originally vacant conduction band of the semiconductor. 

At first, we should note about usual decay channels of photo-excited carriers in the conduction band.
The radiative recombination of an electron-hole pair occurs within $10^{-9}$ second. 
The Auger recombination of an electron-hole pair with no energy dissipation may also 
occur within $10^{-12}$ second or so. 
On the other hand, 
various quantum fluctuations of the electron momentum, 
charge and spin give no energy dissipation. 
Thus, the relaxation phenomenon, we have to describe here, is the intraband one
which occurs in an initial few ps.

Based on above, our scenario is as follows: 
The experimetnt \cite{kanasaki0,kanasaki} has shown that
the electron distributions in the initial stage of energy relaxation
are completely different from the Fermi-Dirac distribution, 
and we cannot define the electron temperature. 
This fact means that the electron-electron scattering is rather rare event 
in the present carrier density  ($\sim 0.003$ electrons per unit cell).
Moreover, the intra-band coulombic scattering among electrons, being completely elastic,
can give no net energy relaxation.
Then, the elastic electron-electron scattering would not be a main relaxation channel 
to reach the Fermi degeneracy at least in the initial stage of the relaxation. 
In the final stage of the energy relaxation, on the other hand,
the electron-distribution becomes close to the Fermi-Dirac distribution 
and the electron-electron scattering has a certain role.
This effect is discussed later with the electron temperature approximation (ETA).
The electron-hole pair attractive interaction does not cause any serious effects 
on the net energy relaxation in the energy region where we are concerning with.
This is because the distribution of the electrons and holes are spatially uniform all through the energy relaxation, 
implying the electron-hole pair attractive interaction is almost within the mean field type.
Thus, the energy relaxation will be dominated by the electron-phonon scattering. 
At the very beginning, the relaxation would be quite rapid, 
since the one-body states of electron below the Franck-Condon one are all completely vacant. 
As the Fermi degeneracy is approached, however, it slows down infinitely, 
since only low energy phonons are available. 
To examine this scenario, we consider a system, 
in which many electrons are coupled with acoustic (ac) and optical (op) phonons.
Our Hamiltonian($\equiv H$) is given as follows:
%
%
\beq
H \equiv  H_{\rm{\scriptsize 0}}+H_{\rm{\scriptsize I}},   
\label{hall}
\eeq
\beq 
H_{\rm{\scriptsize 0}}=\sum_{\bm{k},\sigma} \varepsilon_{\bm k} 
\adag_{{\bm k},\sigma} a_{{\bm k},\sigma}+
\sum_{\bm q} \omega_{\bm q} \bdag_{\bm q} b_{\bm q} 
+ \omega_{\mbox{\scriptsize op}} \sum_{\bm q} \fdag_{\bm q} f_{\bm q}, 
\label{h0}
\eeq 
\beq
H_{\rm{\scriptsize I}} =\frac{S}{\sqrt{2N}} \sum_{{\bm q},{\bm k},\sigma}
\left( \bdag_{\bm q}+b_{\bm{ -q}} \right) \adag_{\bm{ k-q},\sigma} a_{{\bm k},\sigma}  
 +\frac{1}{\sqrt{N}} \sum_{\bm{q},\bm{k},\sigma}
 \frac{iV}{\bm q}\left(
 \fdag_{\bm q} \adag_{\bm{ k-q}, \sigma} a_{{\bm k},\sigma}
- f_{\bm q} \adag_{{\bm k},\sigma} a_{\bm{ k-q},\sigma}
\right). 
\label{hi}
\eeq
Here, $\adag_{{\bm k},\sigma}(a_{{\bm k},\sigma})$ is creation (annihilation) operator of 
an electron with wavevector ${\bm k}$ and spin $\sigma$.
The conduction band energy is assumed to be isotropic and parabolic: 
$\varepsilon_{\bm k} =B\cdot (\bm{ k \cdot k}/\pi^2)$ with 
a band width $B$.
$\bdag_{\bm q}(b_{\bm q})$ and $\fdag_{\bm q}(f_{\bm q})$ 
are creation (annihilation) operators of 
ac and op phonons, respectively, with a wavevector ${\bm q}$.
The acoustic phonon energy is linearly ${\bm q}$-dependent : 
$\omega_{\bm q} = \omega_{\mbox{\scriptsize M}} \cdot ({\bm q}/\pi)$, 
while the op one is ${\bm q}$-independent. $S$ and $V$ are electron-ac and -op phonon coupling constants, respectively.
The second term of $H_{\rm{\scriptsize I}}$ describes the electron-op phonon interaction introduced in Ref. \cite{llp}.
$N$ is the total number of lattice site.

The density matrix $(\equiv\rho(t))$ at a time $t$ is written as a direct product of
the electron density matrix $(\equiv\rho_{\rm{\scriptsize e}} (t))$ and the phonon one 
$\rho_{\rm{\scriptsize p}} \equiv \exp{(-H_{\rm{\scriptsize p}}/k_{\rm{\scriptsize B}} T_{\rm{\scriptsize p}} )}$
with a phonon temperature $T_{\rm{\scriptsize p}}$ .
The phonon system always behaves as a heat reservoir in the present scenario,
and thus $T_{\rm{\scriptsize p}}=0$ K.
An expectation value of electron population at a site ${\bm l}$ is given as 
$\langle n_{{\bm l},\sigma}(t) \rangle \equiv 
{\rm Tr} [ n_{ {\bm l},\sigma } \rho(t)/\rm{Tr} (\rho(t)) ] $,
where $n_{{\bm l},\sigma}\equiv \adag_{{\bm l},\sigma}a_{{\bm l},\sigma}$, and
$a_{ {\bm l},\sigma}\equiv N^{-1/2}\sum_k e^{i{\bm k} \cdot {\bm l}} a_{{\bm k}, \sigma}$.
It is easily shown that $<n_{{\bm l}, \sigma}(t)>$ is independent of ${\bm l}$ and $t$, and
we put $n_{\rm{\scriptsize M}}\equiv <n_{{\bm l}, \sigma}(t)>$.

\begin{figure}[tbp]
\centering
\includegraphics[height=8.0cm,width=7.5cm]{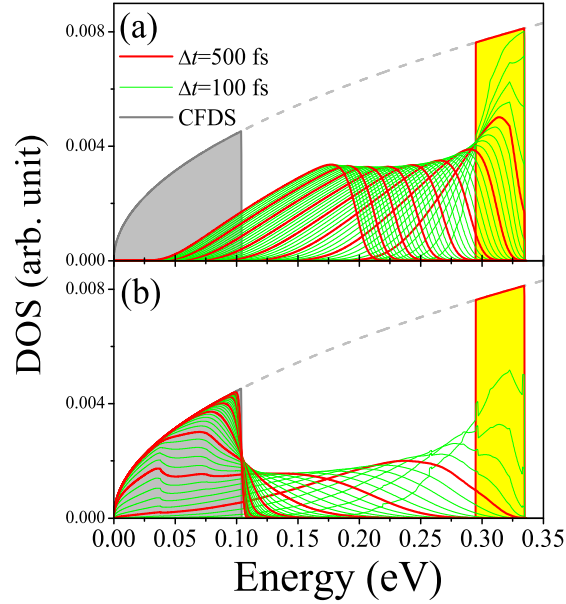} 
\caption{(Color online) Time evolution of DOS is shown.
CFDS means completely fermi degenerated state.
Thick and thin lines are plotted at every 500 and 100 fs, respectively, 
from $t=0$ fs to 4000 fs.}
\label{fig1}
\end{figure}

Our electronic system is in a non-equilibrium state starting from the photo-excitation.
Time evolution of the density matrix is obtained by 
\beq
\tilde{\rho}(t+\Delta t)=\exp_+ \left\{ i\int^{\Delta t}_0 d\tau \tilde{H}_{\rm{\scriptsize I}}(\tau) \right\}
\rho_e(t)\rho_p  
\exp_- \left\{ -i\int^{\Delta t}_0 d\tau' \tilde{H}_{\rm{\scriptsize I}}(\tau') \right\},
\label{rhodef}
\eeq
where subscript $\pm$ in exponent means chronological order and $\tilde{O}(\tau) \equiv e^{i\tau H_{\rm{\scriptsize 0}}} O e^{-i\tau H_{\rm{\scriptsize 0}}} $. 
By expanding the exponential terms in Eq. (\ref{rhodef}) 
up to the second order of $H_{\rm{\scriptsize I}}$,
we can evaluate a time evolution of $\langle n_{\bm{k},\sigma}(t) \rangle$. 
After a tractable calculation with a use of the Fermi's golden rule, 
we get the rate equation for the electron population as
\beq
\frac{\partial n_{{\bm k},\sigma}}{\partial t }=(1- \langle n_{{\bm k},\sigma}(t) \rangle )(
\Gamma^{+(\mbox{\scriptsize ac})}_{\mbox{\scriptsize ep},{\bm k}} (t) +
\Gamma^{+(\mbox{\scriptsize op})}_{\mbox{\scriptsize ep},{\bm k}} (t) ) 
- \langle n_{{\bm k},\sigma}(t) \rangle (
\Gamma^{-(\mbox{\scriptsize ac})}_{\mbox{\scriptsize ep},{\bm k}} (t) +
\Gamma^{-(\mbox{\scriptsize op})}_{\mbox{\scriptsize ep},{\bm k}} (t)
),
\label{rateeq}
\eeq
where
\begin{eqnarray}
\Gamma^{+(\mbox{\scriptsize i})}_{\mbox{\scriptsize ep},{\bm k}}
=C_{\mbox{\scriptsize i}} \sum_{\bm q} \langle n_{ \bm{k+q},\sigma} \rangle  
\delta(\varepsilon_{\bm k} +\omega_{\bm q} -\varepsilon_{\bm{k+q}}), \\
\Gamma^{-(\mbox{\scriptsize i})}_{\mbox{\scriptsize ep},{\bm k}}
=C_{\mbox{\scriptsize i}} \sum_{\bm q} (1- \langle n_{\bm{k+q},\sigma} \rangle ) 
\delta(\varepsilon_{\bm{k+q}} +\omega_{\bm q} -\varepsilon_{{\bm k}}),
\label{gamp}
\end{eqnarray}
with $C_{\mbox{\scriptsize i}}=\pi S^2N^{-1}$ or $ 2\pi V^2N^{-1}$ for i=ac or op, respectively.
Actual numerical calculations are performed by replacing the discretized $\bm{q}$ 
to continuous one, as 
$N^{-1}\sum_{\bm{q}} \rightarrow (4\pi^4/3)^{-1}\int d{\bm{q}}$.
It should be noted that, in our model, the electronic system is relaxed 
all through the process by an "infinite repetition" of a phonon emission 
within the second order perturbation.
Then, all the higher order processes, 
which can be reduced to this infinite repetition of this second order process, 
are taken into account. 

We use the following parameters, which are appropriate for GaAs and InP: 
$B=5$ eV, $\omega_{\rm{\scriptsize M}}=24$ meV,
$\omega_{\rm{\scriptsize op}}=38$ meV, $S=0.5$ eV, and $V=0.13$ eV.   
In our model, there is no anisotropy in ${\bm k}$-space.
Then, under the polar coordinate, 
we use equi-spaced $10^4 k$-mesh in $[0:\pi]$ for the radial component,
and the angle dependent part is convoluted.
Thus, the total number of states is given as
$2N=2\cdot \sum_{k}4\pi k^2$, and the total electron number is 
$n_{\rm{\scriptsize tot}}=n_{\rm{\scriptsize M}}N$. 
Time step $\Delta t$ is set to 0.01 fs.
 
\begin{figure}[tbp]
\centering
\includegraphics[height=9.0cm,width=7.5cm]{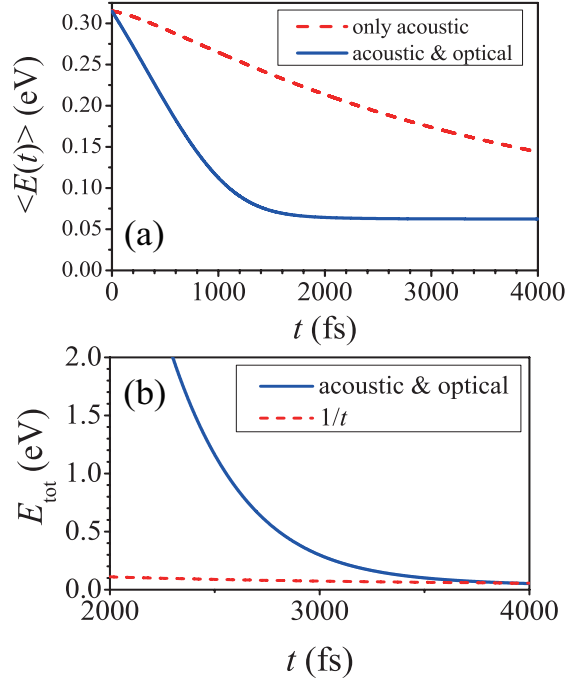} 
\caption{(Color online) Time evolution of (a) the mean energy and (b) total energy are
shown. Definition of these values are given in the text.
In (b), the dashed line represents the energy relaxation which is proportional to $t^{-1}$. 
}
\label{fig2}
\end{figure}

The time evolution of DOS from 0 fs to 4000 fs with $n_{\rm{\scriptsize M}} = 0.003$ 
electrons per site is given in Fig. \ref{fig1}.
We assume that the photo-excited electrons are strongly concentrated around a certain energy,
and the electron population at $t=0$ fs has a rectangular shape.
We have checked even if a gaussian-shaped distribution is adopted for the initial one,
the basic aspect is not altered so much.
When the system is relaxed only by the electron-ac phonon scattering
(Fig. \ref{fig1}a), 
the energy relaxation proceeds very slowly, 
ant its tail can only slightly reach the CBM.
Even after 4000 fs from the photo-excitation,
the state has still not reached around the Fermi degeneracy,
as shown in Fig. \ref{fig2}a.
In Fig. \ref{fig2}a, the mean energy is defined as 
$\langle E \rangle \equiv E_{\rm{\scriptsize tot}}/ n_{\rm{\scriptsize tot}}$,
where $E_{\rm{\scriptsize tot}}\equiv \int E n(E) dE $, which are referenced from the CBM.

When we also take into account the electron-op phonon scattering,
the situation changes drastically (Fig. \ref{fig1}b).
Even in the early stage, a substantial part of electrons relaxes below the Fermi energy very rapidly,
and the broadening of the electron distribution occurs.
This occupation of states below the Fermi energy
is attained by the energy relaxation from higher energy states.
As a result, we can find a multi-peaked distribution, around 1000 fs.
Thus, in this stage, the system shows a quite rapid relaxation (avalanching) as shown in Fig. \ref{fig2}a.

\begin{figure}[tbp]
\centering
\includegraphics[height=5.0cm,width=7.0cm]{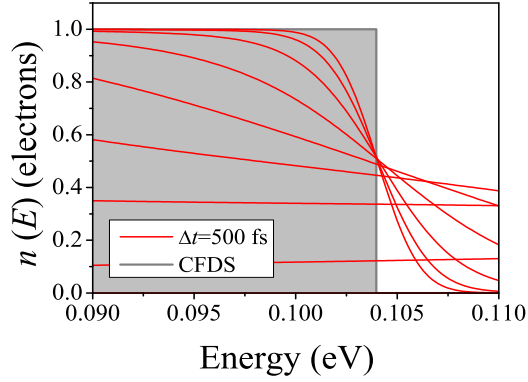} 
\caption{(Color online) Time evolution of electron occupations to 4000 fs is given. 
In CFDS, the electron occupation
is corresponding to Fermi-Dirac distribution function at $T=0$ K.}
\label{fig3}
\end{figure}

After this avalanching phenomenon, 
the energy relaxation shows a critical slowing down at around $t=1200$ fs.
This is because the relaxation occurs only around the Fermi energy,
in which electron scatterings with only low energy phonons are possible.
More precisely, 
the energy of the op phonon is too large for electrons just above the Fermi energy  
to relax down to the unoccupied states just below the Fermi energy.
Thus, the main relaxation process changes from the electron-op phonon scattering to
the electron-ac phonon one.
In Fig. \ref{fig2}a, the shift of the mean energy after the 2000 fs is almost invisible,
but a quite slow relaxation still continues as seen in Fig. \ref{fig2}b,
in which the total energy is referenced from that in the completely Fermi degenerated state (CFDS).
In Fig. \ref{fig3}, the time evolution of the electron distribution is given.
In the later time, the electron distribution approximately follows 
the Fermi-Dirac distribution, implying the electron temperature is well-defined,
while, one can see, it can never be defined in the avalanching process.

The results in Figs. \ref{fig2} and \ref{fig3} lead a conclusion that 
it takes an infinite time to reach the complete Fermi degeneracy since
the energy relaxation continues to slow down.
These results have no serious carrier number dependence,
provided that $n_{\rm{\scriptsize M}}\sim 0.003$.

%
%
The long time limit, or the final stage of the energy relaxation can also be investigated 
by using the ETA.
In the final stage, as mentioned before, the energy relaxation by the electron-ac phonon 
scattering becomes dominant.  
Then, we neglect the electron-op phonon interaction in this stage.
At each time $t$, an electron temperature ($\equiv T_{\rm{\scriptsize e}}$) will be 
always well established in the electronic system, prescribed by 
$H_{\rm{\scriptsize e}}=\sum_{\bm{k},\sigma} \varepsilon_{\bm k} n_{{\bm k},\sigma}$, 
due to intra-system multiple scattering 
by the elastic electron-electron interaction,
though we have not written it explicitly in Eq. (\ref{hall}).
It is weak but becomes effective in the long time limit.
Then, the electron density matrix is given as 
$\rho_{\rm{\scriptsize e}}(t) \equiv \exp{(-H_{\rm{\scriptsize e}}/k_{\rm{\scriptsize B}}
T_{\rm{\scriptsize e}}(t))}$.
$T_{\rm{\scriptsize e}}(t)$ gradually decreases, as releasing its energy to the
phonon system through 
the electron-phonon interaction. 
Thus, we can forget about the electron-electron interaction, except 
$T_{\rm{\scriptsize e}}(t)$.
The electronic heat capacity is defined as
\beq
C(T_{\rm{\scriptsize e}}(t)) \equiv
\frac{\partial \langle H_{\rm{\scriptsize e}} \rangle }{\partial T_{\rm{\scriptsize e}}},
\label{heatcapa}
\eeq
where 
$\langle H_{\rm{\scriptsize e}} \rangle = \sum_{\bm{k},\sigma}(\varepsilon_{\bm{k}}-\mu)
\langle n_{{\bm k},\sigma} \rangle $ with chemical potential $\mu$.
On the other hand, the time evolution of the phonon energy,
$H_{\rm{\scriptsize p}}=\sum_{\bm q} \omega_{\bm q} \bdag_{\bm q} b_{\bm q} $,
through the electron-phonon scattering is calculated by using the aforementioned
density matrix. Along the same manner with the deviation of the rate equation, we obtain
\beq
\langle H_{\rm{\scriptsize p}} (t+\Delta t) \rangle = \Delta t \Gamma (T_{\rm{\scriptsize e}}),
\label{hp}
\eeq
where
\beq
\Gamma (T_{\rm{\scriptsize e}})= \frac{\pi S^2}{N} \sum_{\bm{q,k},\sigma} \omega_{\bm{q}}
(1-\langle n_{\bm{k-q},\sigma} \rangle) \langle n_{\bm{k},\sigma} \rangle 
\delta(\omega_{\bm{q}}+\varepsilon_{\bm{k-q}}-\varepsilon_{\bm{k}}).
\label{hpgam}
\eeq
Then, by considering the energy conservation:
\beq
\Delta \langle H_{\rm{\scriptsize e}}(T_{\rm{\scriptsize e}})\rangle
= C(T_{\rm{\scriptsize e}}) | \Delta T_{\rm{\scriptsize e}} |
= \Gamma (T_{\rm{\scriptsize e}}) \Delta t
= \Delta \langle H_{\rm{\scriptsize p}}(T_{\rm{\scriptsize e}}) \rangle,
\label{eneconservation}
\eeq
we get the equation for the 
electronic system cooling as
\beq
\frac{\partial T_{\rm{\scriptsize e}}}{\partial t} =
-\frac{\Gamma(T_{\rm{\scriptsize e}})}{C(T_{\rm{\scriptsize e}})}.
\label{coolingeq}
\eeq
At low temperature, 
the electronic heat capacity is well-known to be linearly proportional to 
temperature \cite{luttinger}. 
Electron-hole pair numbers around the Fermi energy and ac phonon energy 
are also proportional to 
$T_{\rm{\scriptsize e}}$, while the phonon mode density is proportional 
to $T^2_{\rm{\scriptsize e}}$.
Then, $\Gamma(T_{\rm{\scriptsize e}}) \propto T^4_{\rm{\scriptsize e}}$.
From these order estimations with Eq. (\ref{coolingeq}),
we can finally obtain $T_{\rm{\scriptsize e}} \propto t^{-1/2}$,
and thus 
$\Delta \langle H_{\rm{\scriptsize e}}(T_{\rm{\scriptsize e}}) \rangle \propto t^{-1}$.
Although our theory for the electron-ac phonon interaction in Eq.(\ref{hi}) is rather phenomenological,
being a little different from the standard deformation potential one \cite{toyozawa, cardona},
this result agrees well with the experiment \cite{kanasaki0,kanasaki}, and gives
much slower relaxation than the aforementioned results of the real time dynamics,
as shown in Fig. \ref{fig2}b.

In summary, 
the energy relaxation dynamics of the macroscopically photo-excited carriers
in the conduction band is theoretically studied,
by considering the phonon system as a heat reservoir.
The avalanching process soon after the photo-excitation is found, and it is
mainly caused by the electron-op phonon scattering. 
In this process, we found the appearance of an elongated multi-peaked distribution in the DOS.
The slow dynamics is seen in the later time, in which
only the low energy phonon is available, and thus the electron-ac phonon
scattering works as a main reservoir.
At around 1200 fs, main relaxation process is changed from electron-op phonon scattering
to electron-ac phonon one. At that time, the relaxation shows a critical slowing down.
With a use of the ETA for the electronic system cooling,
we showed that the relaxation in the long time  limit is proportional to $t^{-1}$.
The present results well agree with the basic aspects of the recent
time-resolved two-photon photoemission \cite{kanasaki0,kanasaki}.
This conclusion has a close connection with the Luttinger theorem \cite{luttinger2}, and 
the theory for the life time of the quasi-particle in the Fermi liquid \cite{fermiliquid}. 

%
%
\begin{acknowledgment}
The authors would like to thank K. Tanimura and J. Kanasaki for
presenting their research result prior to publication.
This work was partly supported by JSPS Grant-in-Aid 
for Specially Promoted Research, Grant Number 24000006.
\end{acknowledgment}


\end{document}